\documentclass[%
twocolumn,
floatfix,
 reprint,
 amsmath,amssymb,
 aps,
]{revtex4-2}

\usepackage{graphicx}
\usepackage{dcolumn}
\usepackage{bm}
\usepackage{enumerate}   

\usepackage{float}

\preprint{APS/123-QED}

\newcommand{\pp}           {$pp$}

\newcommand{\pPb}          {\mbox{$p$--Pb}}

\newcommand{\s}            {\ensuremath{\sqrt{s}}}
\newcommand{\pt}           {\ensuremath{p_\mathrm{T}}}

\newcommand{\mt}           {\ensuremath{m_{\rm T}}}
\newcommand{\snn}          {\ensuremath{\sqrt{s_{\rm NN}}}}

\newcommand{\gevc}         {\ensuremath{{\rm GeV}/c}}

\newcommand{\pttrig}{\ensuremath{p_\mathrm{T,\,trig}}}
\newcommand{\ptassoc}{\ensuremath{p_\mathrm{T,\,assoc}}}

\begin{document}

\preprint{APS/123-QED}

\title{Toward an unbiased flow measurements in LHC \pp\ collisions}
\author{~S.~Ji$^{1}$} 
\author{~M.~Virta$^{2,3}$} 
\author{~T.~Kallio$^{2}$}
\author{~S.H~Lim$^{1}$}
\author{~D.J.~Kim$^{2,3}$}
\email{su-jeong.ji@cern.ch}
\address{
$^{1}$Pusan National University, Department of Physics, Busan 46241, Republic of Korea\\ 
$^{2}$University of Jyv\"askyl\"a, Department of Physics, P.O. Box 35, FI-40014 University of Jyv\"askyl\"a, Finland\\
$^{3}$Helsinki Institute of Physics, P.O.Box 64, FI-00014 University of
	Helsinki, Finland
 }

\date{\today}

\begin{abstract}
Long-range correlations for pairs of charged particles with two-particle angular correlations are studied in \pp\ at ${\sqrt{{\textit s}}}=13$~TeV with various Monte Carlo generators. The correlation functions are constructed as functions of relative azimuthal angle $\Delta\varphi$ and pseudorapidity separation $\Delta\eta$ for pairs of different particle species with the identified hadrons such as $\pi$, $K$, $p$, and $\Lambda$ in wide $\Delta\eta$ ranges.
Fourier coefficients are extracted for the long-range correlations in several -multiplicity classes using a low-multiplicity template fit method. The method allows to subtract the enhanced away-side jet fragments in high-multiplicity with respect to low-multiplicity events. However, we found that due to a kinematic bias on jets and differing model implementation of flow and jet components, subtracting the non-flow contamination in small systems can bias the results. It is found that PYTHIA8 Default model where the presence of the collective flow is not expected but the bias results in very large flow. Also extracting flow signal from the EPOS4 and PYTHIA8 String Shoving models is not possible because of flow signal introduced in the low-multiplicity events. Only AMPT String Melting model among studied model calculations is free from this bias, and shows a mass ordering at low $p_{\mathrm{T}}$ and particle type grouping in the intermediate $p_{\mathrm{T}}$ range. This feature has first found in large systems but the mass ordering in small systems is different from what is observed in the large collision systems.
\end{abstract}

\maketitle

\section{\label{sec:intro}Introduction}
Collisions between heavy-ions (HIC) exhibit strong collectivity, as demonstrated by the anisotropy in the momentum distribution of final particles emitted at the Relativistic Heavy Ion Collider (RHIC)~\cite{Adams:2005dq,Adcox:2004mh,Arsene:2004fa,Back:2004je} and the Large Hadron Collider (LHC)~\cite{Abelev:2012di, Abelev:2014pua, ATLAS:2011ah}.
The spatial anisotropies are converted to anisotropies in the final momentum distribution
due to a pressure-driven expansion of the strongly interacting quark-gluon plasma (QGP) formed during the collision event.
The produced QGP in HIC is in the strongly coupled regime and the state-of-the-art Bayesian analyses utilizing the experimental data favor small values of the shear viscosity to entropy density ratio ($\eta/s$), which implies that the produced QGP is considered the fluid with the lowest shear viscosity to entropy density ratio observed in nature~\cite{Kovtun:2004de,Bernhard2019}. In Recent years, the primary focus has been to constrain model parameters by measuring sensitive observables, using Bayesian analyses~\cite{Auvinen:2020mpc,Nijs:2020ors,Nijs:2020roc,JETSCAPE:2020mzn,Parkkila:2021tqq,Parkkila:2021yha}.

To probe the collective behavior in the momentum anisotropy, long-range particle correlations are used over a wide range of pseudorapidity. Over the past few years, long-range correlations have also been observed in smaller collision systems, such as high-multiplicity (HM) proton-proton (\pp) collisions~\cite{Aad:2015gqa,Khachatryan:2015lva,Khachatryan:2016txc,Acharya:2019vdf,ATLAS:2017rtr}, proton-nucleus ($p$A) collisions~\cite{ALICE:2012eyl,Aad:2014lta,Aaboud:2016yar,Khachatryan:2016ibd}, and collisions of light ions with heavy ions, such as p+Au, d+Au, $^3$He+Au~\cite{PHENIX:2018lia,Aidala:2017ajz}. These observations raise the question of whether small system collisions have a similar underlying mechanism for developing correlations as heavy AA collisions.

On the experimental side, extracting flow in small systems remains challenging due to a strong jet fragmentation bias to the long-range correlations. One commonly used approach for suppressing the non-flow contribution in two-particle correlations is to require a large $\Delta\eta$ gap between the two particles, which is also applied in cumulant methods~\cite{Bilandzic:2010jr, Acharya:2019vdf}. However, this approach only eliminates non-flow contributions on the near side, not on the away side ($\Delta\varphi\sim\pi$). To address this limitation, a low-multiplicity template fit (LMTF) method has been proposed to remove away-side contributions as well~\cite{ATLAS:2015hzw,ATLAS:2016yzd,ATLAS:2018ngv}, taking into account the autocorrelation between event multiplicity and jet yields~\cite{CMS:2013ycn}. This method enables the subtraction of enhanced away-side jet yields in HM events compared to low-multiplicity (LM) events, and may potentially provide a lower limit on the event multiplicity needed to observe the flow signal.

The observed number of constituent quark (NCQ) scaling pattern of the elliptic flow at RHIC~\cite{STAR:2003wqp,STAR:2007afq,PHENIX:2003qra,PHENIX:2006dpn} and LHC~\cite{ALICE:2014wao,ALICE:2015lib,ALICE:2016cti,ALICE:2018yph} in large collision systems often refers to evidence of the creation of a thermalized bulk system of quarks that coalesce into hadrons. Whether these patterns can still be observed in collisions of small systems is a question of great current interest. The observation of NCQ scaling in smaller systems would provide important insights into the underlying physics of the system.
An approximate NCQ scaling of charged hadrons' $v_2$ in \pPb\ collisions at \snn\ = 5.02~TeV is observed at intermediate $p_\mathrm{T}$ with ALICE ~\cite{Pacik:2018gix} and also for $v_{2}$ of $\pi$ and $p$ in $^3$He+Au collisions at \snn\ = 200 GeV with PHENIX~\cite{PHENIX:2017djs}. However, this observation was based on a limited range of $p_\mathrm{T}$ with the  cumulant methods and further experimental checks are needed to confirm the presence of NCQ scaling over a wider range of $p_\mathrm{T}$ with the experimental LMTF method. Additionally, it is important to note that other effects, such as initial-state fluctuations and final-state correlations, can also contribute to the observed elliptic flow in small systems. Therefore, more detailed studies are needed to understand the interplay of these effects and the possible mechanisms underlying the observed NCQ scaling patterns.

On the theoretical side, systematic mapping of the multiparticle correlations across collision systems by varying sizes is presently underway~(see e.g. \cite{Schenke:2020mbo}). The quantitative description of the full set of experimental data has not been achieved yet. A summary of various explanations for the observed correlations in small systems is given in~\cite{Strickland:2018exs,Loizides:2016tew,Nagle:2018nvi}.

Another important piece of evidence for a strongly interacting medium in small collision systems would be the presence of jet quenching~\cite{Gyulassy:1990ye,Wang:1991xy}. However, no evidence of jet quenching has been observed in either HM \pp\ or \pPb\ collisions~\cite{Adam:2014qja,Khachatryan:2016odn,Adam:2016jfp,Adam:2016dau,Acharya:2017okq}. A study with two-particle angular correlations in short-range correlations around $(\Delta\eta$, $\Delta\varphi)=(0,0)$ is a good tool for studying jet fragmentations~\cite{Adam:2016tsv}.

This report investigates the relationship between jet production and collective phenomena in small systems using various Monte Carlo event generators, such as AMPT~\cite{Lin:2004en}, PYTHIA8 String Shoving~\cite{Bierlich:2017vhg,Bierlich:2019ixq}, and EPOS4~\cite{Pierog:2013ria}. Although all three models incorporate both jets and collective flow effects, they differ in their approach to describing collective flow. To determine the suitability of each model for a specific experimental method, we assess the latest flow extraction technique, LMTF, against these models. 
This paper is organized as follows. First, the model descriptions are given in Sec.~\ref{sec:models} and analysis methods are described in Sec.~\ref{sec:ana}. The results from model calculations are presented in Sec.~\ref{sec:results}. Finally, the results are summarized in Sec.~\ref{sec:summary}.

 \section{\label{sec:models}Model descriptions}
In this study, several Monte Carlo (MC) event generators, such as PYTHIA8, AMPT, and EPOS4, of different characteristics are used to compare the non-flow subtraction results. 
We generate a few million \pp\ collision events with each event generator and collect final-state charged particles for further analysis. Here we have a brief description of the event generators.

\textit{PYTHIA8:}
PYTHIA8 is a widely used event generator for high-energy \pp\ collisions, and it recently incorporates a capability of heavy-ion collisions.
It includes both hard and soft interactions for jets and underlying events, and the default parameter set called Monash tune can reasonably describe the production of soft particles~\cite{Skands:2014pea}.
In the default version, there is no partonic or hadronic interaction, so we do not expect a long-range correlation among produced particles due to the flow contribution.
Hence, it has been used to verify methods to estimate the non-flow contribution~\cite{Lim:2019cys}.

\textit{PYTHIA8 String Shoving:}
In PYTHIA8, a model to describe the long-range correlation in HM \pp\ collisions called ``string shoving'' has been implemented as an option~\cite{Bierlich:2017vhg,Bierlich:2019ixq}. This model introduces a repulsive force between strings, and the interaction can cause a microscopic transverse pressure, giving rise to the long-range correlations. The string shoving approach in PYTHIA8 successively reproduces the experimental measurements of the long-range near-side ($\Delta\varphi\sim0$) ridge yield in HM \pp\ events by ALICE~\cite{ALICE:2021nir} and CMS~\cite{Khachatryan:2016txc}. 
However, strings produced from hard scatterings are also affected by the repulsive force, which then leads to observed long-range correlation even in low-multiplicity events~\cite{Kim:2021elv}. 


\textit{AMPT:}
Besides several models based on the causal hydrodynamic framework in describing the collective evolution in small collision systems, the AMPT model with string melting~\cite{Lin:2004en} can reproduce the flow-like signals by modeling the evolution of medium as a collection of interacting quarks and hadrons~\cite{OrjuelaKoop:2015jss}.
The applicability of fluid-dynamical simulations and partonic cascade models in small systems has been explored in Ref.~\cite{Gallmeister:2018mcn}. In the context of kinetic theory with isotropization-time approximation, the model can smoothly explain the long-range correlations by fluid-like (hydrodynamic) excitations for Pb--Pb collisions and particle-like (or non-hydrodynamic) excitations for \pp\ or \pPb\ collisions~\cite{Kurkela:2019kip,Kurkela:2020wwb,Ambrus:2021fej}. 


\textit{EPOS4:}
The EPOS model describe the full evolution of medium produced by heavy-ion collisions with two parts called a core and a corona~\cite{Werner:2013tya}. 
The core part follows the hydrodynamic expansion, and the corona part is composed of hadrons from string decays.
After the hadronization process of the core part, the UrQMD model is used to describe hadronic interactions among all hadrons from two parts.   
The version called EPOS LHC including a different type of radial flow in the case of a small but a very dense system can successfully describe the long-range correlation in HM \pp\ events~\cite{ALICE:2021nir}. 
Recently, a new version of EPOS (EPOS4) has been released to the public.
We utilize the framework for this study.

\begin{table}[h!]
\resizebox{0.45\textwidth}{!}{
\begin{tabular}{|c|c|c|}
\hline
 Models & Characteristics & Mechanism \\ 
 \hline
PYTHIA8 Default     & jets and no flow &  Ref.~\cite{Skands:2014pea} \\ 
 \hline
PYTHIA8 Shoving    & jets and flow &  String repulsion~\cite{Bierlich:2017vhg,Bierlich:2019ixq} \\ 
 \hline
AMPT      & jets and flow &  String melting\cite{Lin:2004en} \\
\hline
EPOS4     & jets and Hydro & Core (hydrodynamical)~\cite{Pierog:2013ria}  \\
\hline
 \end{tabular}
 }
 \caption{A list of the models used in this paper.}
 \label{tab:models}
 \end{table}

 The summary of the model characteristics is listed in Tab.~\ref{tab:models}. The PYTHIA8 Default model is used to understand the non-flow contributions. 
The PYTHIA8 Shoving, AMPT, and EPOS4 models all include both jets and collective flow effects. However, they differ in their mechanisms for describing the collective flow. It is important to note that the applicability of each model to a specific experimental method may depend on various factors, such as the collision system being studied, as well as the specific observables being measured. Therefore, it is important to carefully consider the strengths and limitations of each model when interpreting experimental results. For instance, in the study by ALICE~\cite{ALICE:2021nir}, both PYTHIA8 Shoving and EPOS4 fail to reproduce the near-side jet yields, with PYTHIA8 Shoving predicting an increasing near-side jet yield with increasing multiplicity, while EPOS4 shows the opposite trend. Regarding the ridge yields, EPOS4 overestimates them, while PYTHIA8 Shoving underestimates them. The ridge yields in low multiplicity events are similar to those in HM events for EPOS4 and PYTHIA8 Shoving, while they decrease towards low multiplicity events in the experimental data~\cite{CMS:2015fgy}.

\section{Analysis procedure}
\label{sec:ana}
\subsection{Event and particle selections}

\begin{figure}[h!]
		\includegraphics[width=0.99 \columnwidth]{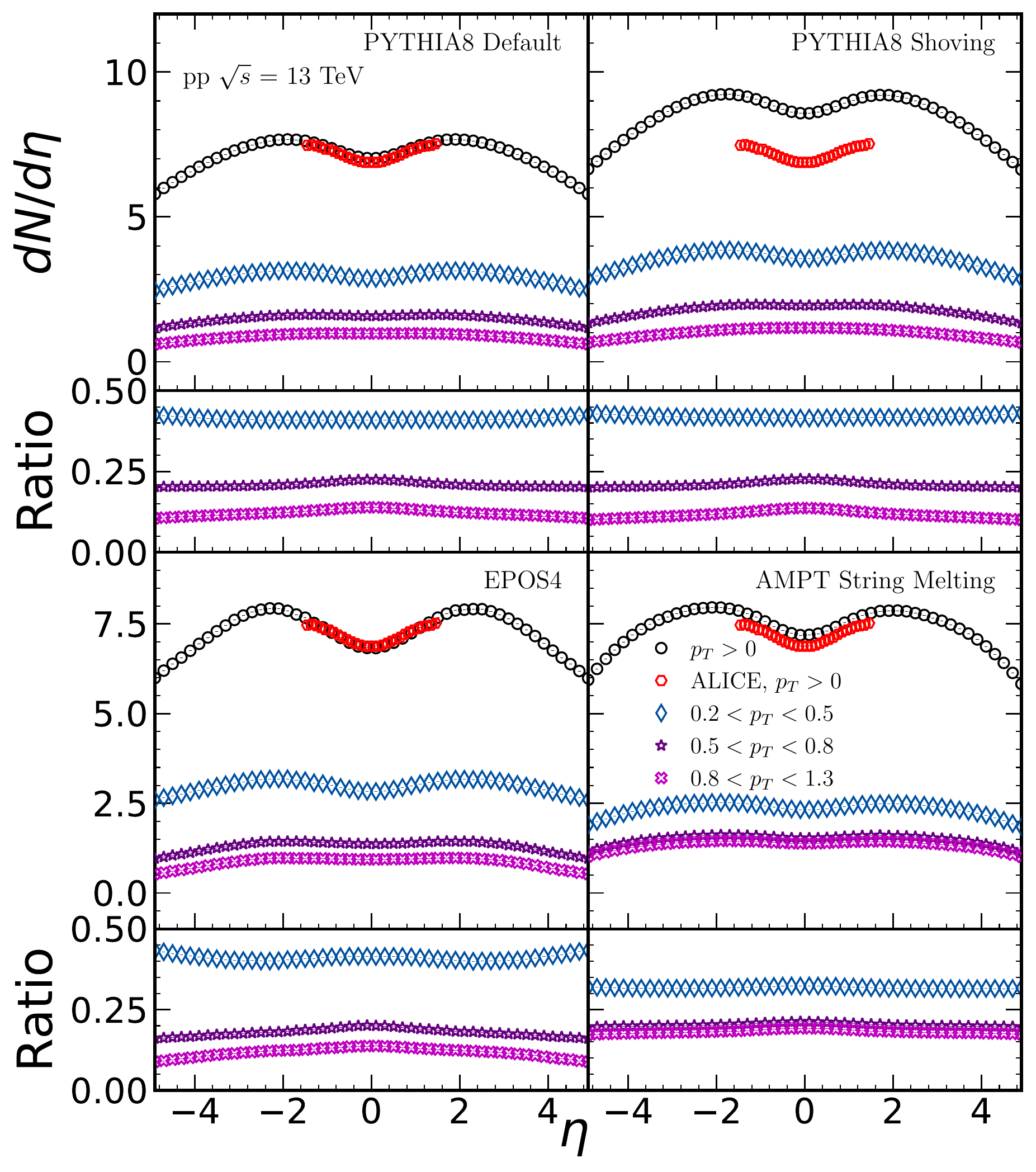} 
\caption{Charged--particle pseudorapidity density for four different \pt\ intervals over a broad $\eta$ ranges in several model calculations is compared to the ALICE data~\cite{ALICE:2020swj}.}
\label{fig:dndeta}
\end{figure}

This analysis uses the same event selection criteria as the ALICE experiments, which require a charged particle in both V0A and V0C~\cite{ALICE:2008ngc,ALICE:2013axi} acceptance. The V0A and V0C cover the pseudorapidity ranges $2.8 < \eta < 5.1$ and $-3.7 < \eta < -1.7$, respectively. The contribution from diffractive interactions is minimized in these events~\cite{ALICE:2020swj}.
Fig.~\ref{fig:dndeta} shows the charged particle density in various $\pt$ intervals. 
Every model describes the trend of the data well, while PYTHIA8 String Shoving and AMPT model overestimates the data from the ALICE collaborations~\cite{ALICE:2020swj}.
Despite of PYTHIA8 String Shoving model largely overestimates the data, the $\pt$ dependency is similar with PYTHIA8 Default and EPOS4.
In the case of the AMPT model, it shows the different $\pt$ dependency.

\begin{figure}[h!]
		\includegraphics[width=0.99 \columnwidth]{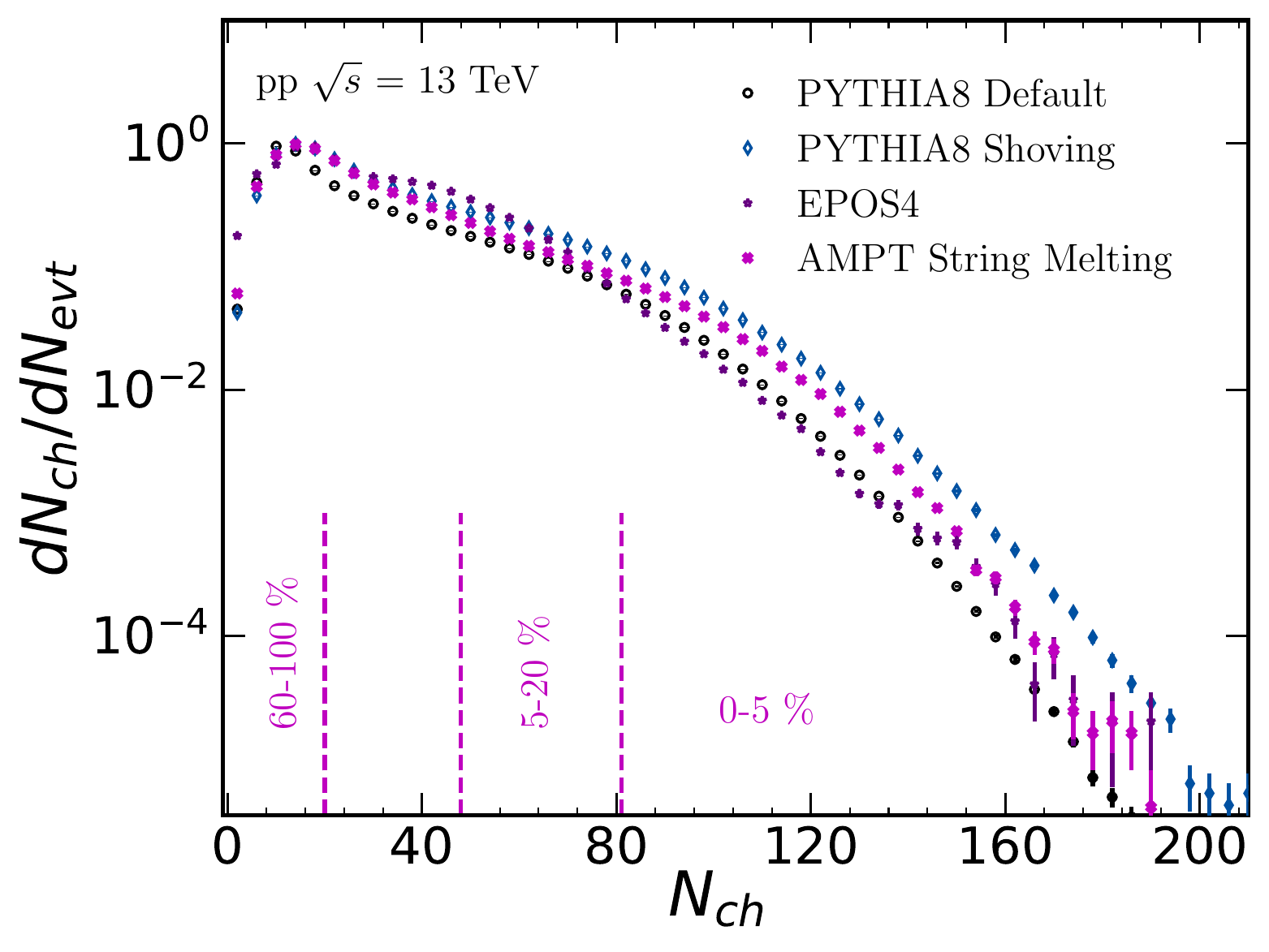} 
\caption{The distribution of the V0M charged particles in the region $-3.7<\eta<-1.7$ and $2.8<\eta<5.1$. This is used to determine the event multiplicity classes in \pp\ collisions at $\s$~=~13 TeV.}
\label{fig:mult}
\end{figure}

The multiplicity percentiles are estimated by V0M, which is the sum of the charged particles both in the V0A and V0C acceptance.
The event multiplicity of V0M from different generators is shown in Fig.~\ref{fig:mult}.
PYTHIA8 String Shoving model generates HM events more than other models.
The vertical lines indicates the 0--5\%, 5--20\% and 60--100\% event multiplicity of AMPT String Melting events.
For the identified flow measurement, $\pi$, $K$, and $p$ for all models and additionally $\Lambda$ for AMPT model are studied by selecting the particle identification code from the models in the range of $0.2<p_\mathrm{T}<6~\gevc$.

\subsection{Two-particle angular correlations}
Two-particle angular correlations are measured as functions of the relative azimuthal angle ($\Delta\varphi$) and the relative pseudorapidity ($\Delta\eta$) between a trigger and associated particles
\begin{eqnarray}
\frac{1}{N_{\rm{trig}}} \frac{ \rm{d}\it{}^{2} N_{\rm{pair}} }{ \rm{d} \Delta\eta \rm{d}\Delta\varphi} = B_{max}\frac{S(\Delta\eta, \Delta\varphi)}{B(\Delta\eta, \Delta\varphi)}  \Big\lvert_{\pttrig,\,\ptassoc}\,,
\label{eq:corrfunction}
\end{eqnarray}
where the trigger and associated particles are defined for different transverse momentum ranges and different $\eta$ acceptance of the detectors.
The $N_\mathrm{trig}$ and $N_\mathrm{pair}$ are the numbers of trigger particles and trigger-associated particle pairs, respectively. $S(\Delta\eta, \Delta\varphi)$ corresponds to the average number of pairs in the same event and $B(\Delta\eta, \Delta\varphi)$ to the number of pairs in mixed events. $B_{max}$ represents the normalization of $B(\Delta\eta, \Delta\varphi)$, and by dividing $S(\Delta\eta, \Delta\varphi)$ with $B(\Delta\eta, \Delta\varphi)/B_{max}$ the acceptance effects are corrected for.
This analysis is performed for several multiplicity percentiles (0--5\%, 0--20\%, 20--40\%, and 60--100\%), and for each multiplicity percentile. 

The flow studies using the ALICE detector were carried out using only the particles detected in the TPC detector~\cite{ALICE:2020swj}. However, due to the limited $\eta$ acceptance of the TPC detector, the study was restricted to the edge of the detector with $1.6<\Delta\eta|<1.8$, as well as $p_{\mathrm{T}}>1.0$ GeV/$c$ to avoid non-flow contributions~\cite{ALICE:2020swj}.
To further suppress non-flow contributions, preliminary studies by the ALICE experiment have used the very forward FMD detectors to achieve a large $\eta$ separation of the correlated particles, up to $|\Delta\eta| \approx 6$. In this analysis, we use the same combinations of correlations between particles in the TPC and FMD detectors.

Tab.~\ref{tab:acc} lists the $\eta$ acceptance and measurable $p_{\mathrm{T}}$ ranges for each detector used in the analysis.

\begin{table}[h!]
\begin{tabular}{|c|c|c|}
\hline
 Detector & $\eta$ acceptance & $\pt$ range \\ 
 \hline
 TPC      & $|\eta| < 0.8$ & $0.2< \pt < 6.0~\gevc$ \\ 
 \hline
 FMDA      & $1.9<\eta<4.8$ & $\pt >0.0~\gevc$ \\ 
 \hline
 FMDC      & $-3.1 <\eta <-1.9$ & $\pt >0.0~\gevc$  \\
\hline
 \end{tabular}
 \caption{The acceptance of the detectors used for the trigger and/or associated particles.}
 \label{tab:acc}
 \end{table}

\begin{figure*}[tbh!]
	\centering
	\includegraphics[width=1.0\linewidth]{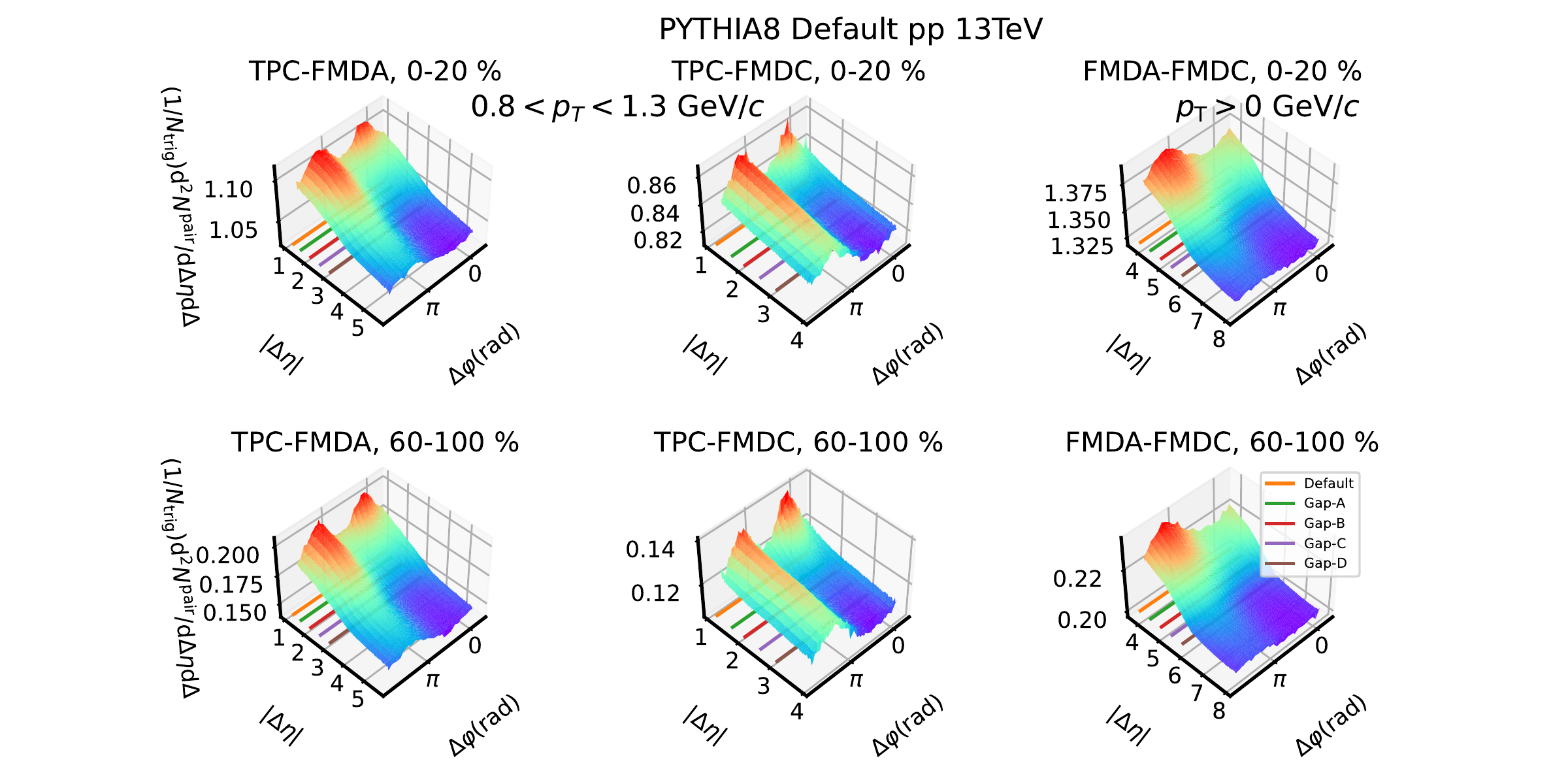} 
	\caption{Two-particle correlation functions as functions of $\Delta\eta$ and $\Delta\varphi$ for HM (0--20\%, top panels) and LM (60--100\%, bottom panels) events using different combinations of the detectors in \s\ = 13 TeV \pp\ collisions from AMPT String Melting calculations. The intervals of $\pttrig$ and $\ptassoc$ are 0.8~$<\it{p}_{\rm{T}}<$~1.3~GeV/$c$ with TPC and $\it{p}_{\rm{T}}>$~0~GeV/$c$ with FMDA or FMDC.}
	\label{fig:2DPythiaDefault}
\end{figure*}

As for TPC–FMD correlations, the trigger particles are from TPC detectors with various $\pt$ intervals and the associated particles are from FMDA or FMDC in a different $\eta$ ranges with $\pt >0.0~\gevc$. As for FMDA–FMDC correlations, both trigger and associated particles come from FMD detector with $\pt >0.0~\gevc$. The $\Delta\eta$ ranges used for the default analysis with the full $\eta$ acceptance of all detectors and four additional wider $\Delta\eta$ gaps used further to reduce the non-flow contributions are summarized in Tab.~\ref{tab:accDeta}.
\begin{table}[htb]
\resizebox{0.45\textwidth}{!}{
\begin{tabular}{|c|c|c|c|c|c|}
\hline
 Correlations & Default &  Gap-A &  Gap-B &  Gap-C &  Gap-D\\ 
 \hline
 TPC-FMDA      & [1.1, 5.6] & [1.5, 5.6] & [2.0, 5.6] & [2.5, 5.6] & [3.0, 5.6]\\ 
 \hline
 TPC-FMDC      & [1.1, 3.9] & [1.6, 3.9] & [2.0, 3.9] & [2.5, 3.9] & [3.0, 3.9]\\ 
 \hline
 FMDA-FMDC      & [3.8, 7.9] & [4.3, 7.9]  & [4.8, 7.9] & [5.3, 7.9] & [5.8, 7.9]\\
\hline
 \end{tabular}
 }
 \caption{The $|\Delta\eta|$ ranges of each correlation function and four additional wider $\Delta\eta$ gaps used further to reduce the non-flow contributions.}
 \label{tab:accDeta}
 \end{table}

\begin{figure*}[tbh!]
	\centering
	\includegraphics[width=1.0\linewidth]{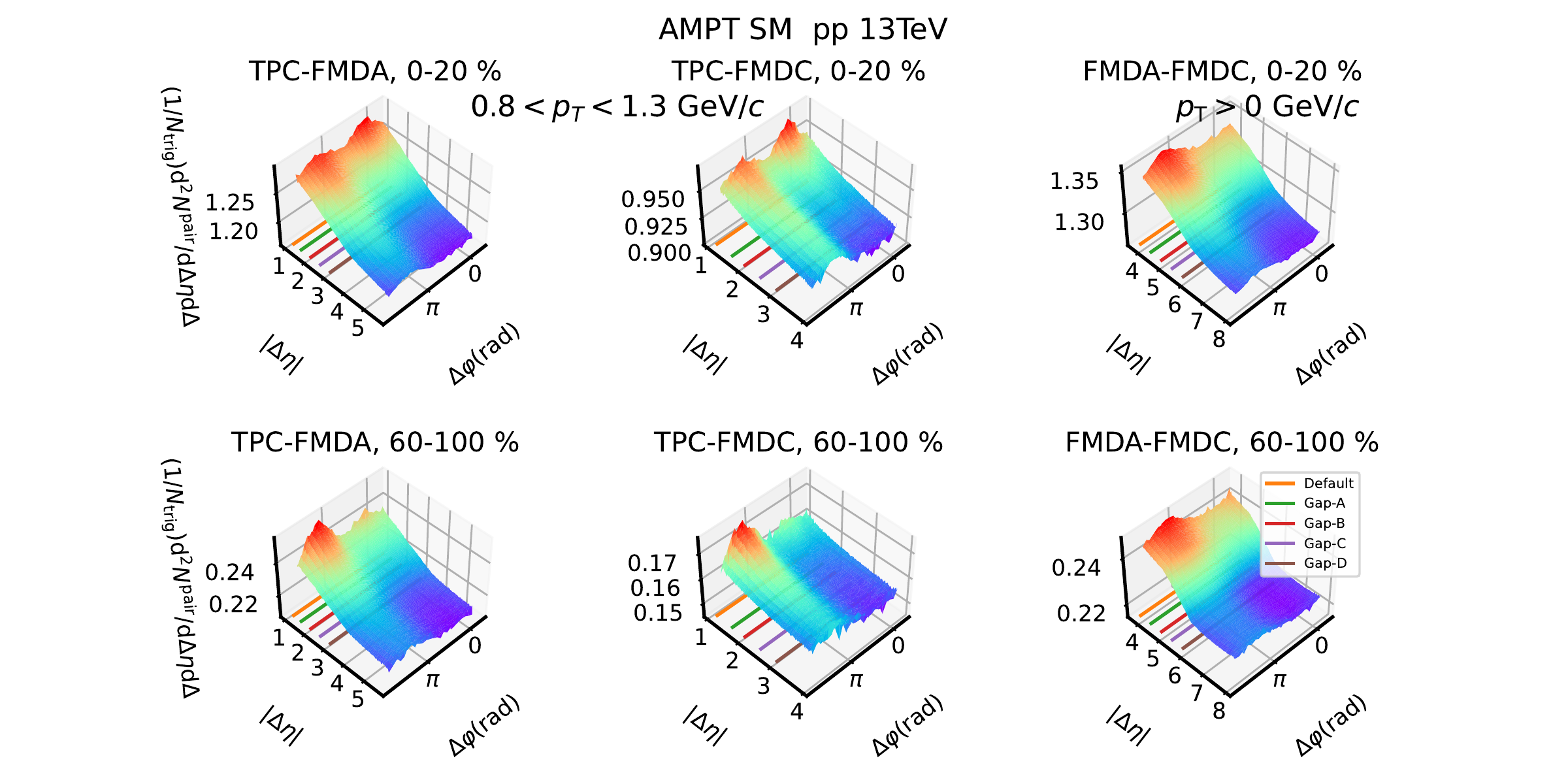} 
	\caption{Two-particle correlation functions as functions of $\Delta\eta$ and $\Delta\varphi$ for HM (0--20\%, top panels) and LM (60--100\%, bottom panels) events using different combinations of the detectors in \s\ = 13 TeV \pp\ collisions from PYTHIA8 Default calculations. The intervals of $\pttrig$ and $\ptassoc$ are 0.8~$<\it{p}_{\rm{T}}<$~1.3~GeV/$c$ with TPC and $\it{p}_{\rm{T}}>$~0~GeV/$c$ with FMDA or FMDC. (for PYTHIA8 string shoving see the Appendix Fig. 15)}
	\label{fig:2DAMPT_SM}
\end{figure*}

Fig.~\ref{fig:2DPythiaDefault} and Fig.~\ref{fig:2DAMPT_SM} show the 2-dimensional correlation function of each detector combination with the events from PYTHIA8 Default and AMPT String Melting models, respectively. Unlike the events from AMPT having both flow and jet components, the PYTHIA8 Default events contain the particles purely from jets. The peak seen in the short-range represents the jet contribution. Even though already having long-range correlations by using the particles in TPC and FMD, still the large jet contamination is seen. To find a safe long-range region for the analysis, five different long-ranges are selected to study the effect on the degree of the jet contamination. Different shape and the amplitude of the jet peak is seen depending on the models.

In the next section, the details about the LMTF method, which is used for the non-flow subtraction, will be discussed as well as the  assumptions of the method.

\subsection{Extraction of flow coefficients from the Low-Multiplicity Template Fit Method}
Due to the strong jet fragmentation bias in small collision systems it is difficult to extract the flow in these collisions because of the remaining non-flow in the away-side region ($\Delta\varphi \sim \pi$) in Eq.~\ref{eq:corrfunction}. As discussed in Refs.~\cite{ATLAS:2015hzw,ATLAS:2016yzd}, the HM correlation function in a HM percentile can be expressed as 
\begin{equation}
\label{eq:1}
\begin{aligned}
Y_{\rm{HM}}(\Delta\varphi) = G~(1 + 2v_{2,2}\cos(2\Delta\varphi) \\
+ 2v_{3,3}\cos(3\Delta\varphi) \\
+ 2v_{4,4}\cos(4\Delta\varphi))  \\
+ F~Y_{\rm{LM}}(\Delta\varphi) \quad,
\end{aligned}
\end{equation}
where $Y_{\rm{LM}}(\Delta\varphi)$ is the LM correlation function, G is the normalization factor for the Fourier component up to the fourth harmonic, and the scale factor $F$ corresponds to the relative away-side jet-like contribution with respect to the low-multiplicity (LM) (the 60--100\%).
This method assumes that $Y_{\rm{LM}}$ does not  contain a peak in the near side originating from jet fragmentation and that the jet shape remains unchanged in HM events compared to LM events.
The first assumption is well-verified using the selected LMTF for the experimental data~\cite{ATLAS:2018ngv}, while the second assumption regarding the modification of jet shapes is tested using the near-side $\Delta\eta$ distributions. Additionally, the ATLAS Collaboration's study of HM \pp\ and \pPb\ collisions in Ref.~\cite{ATLAS:2018ngv} provides further support for this assumption, as there is no evidence of jet quenching in these collisions~\cite{Adam:2014qja,Khachatryan:2016odn,Adam:2016jfp,Adam:2016dau,Acharya:2017okq}.
The fit determines the scale factor $F$ and pedestal $G$, and $v_{n,n}$ are calculated from a Fourier transform. It is worthwhile noting that this method does not rely on the zero yield at minimum (ZYAM) hypothesis to subtract an assumed flat combinatorial component from the LMTF as done previously in Refs.~\cite{ATLAS:2012cix,ATLAS:2014qaj}. Whether or not if the models agree on the assumption about the jet shape modification depending on the event multiplicity will be discussed in the Sec.~\ref{sec:results}.

\begin{figure}[h!]
	\includegraphics[width=0.99 \columnwidth]{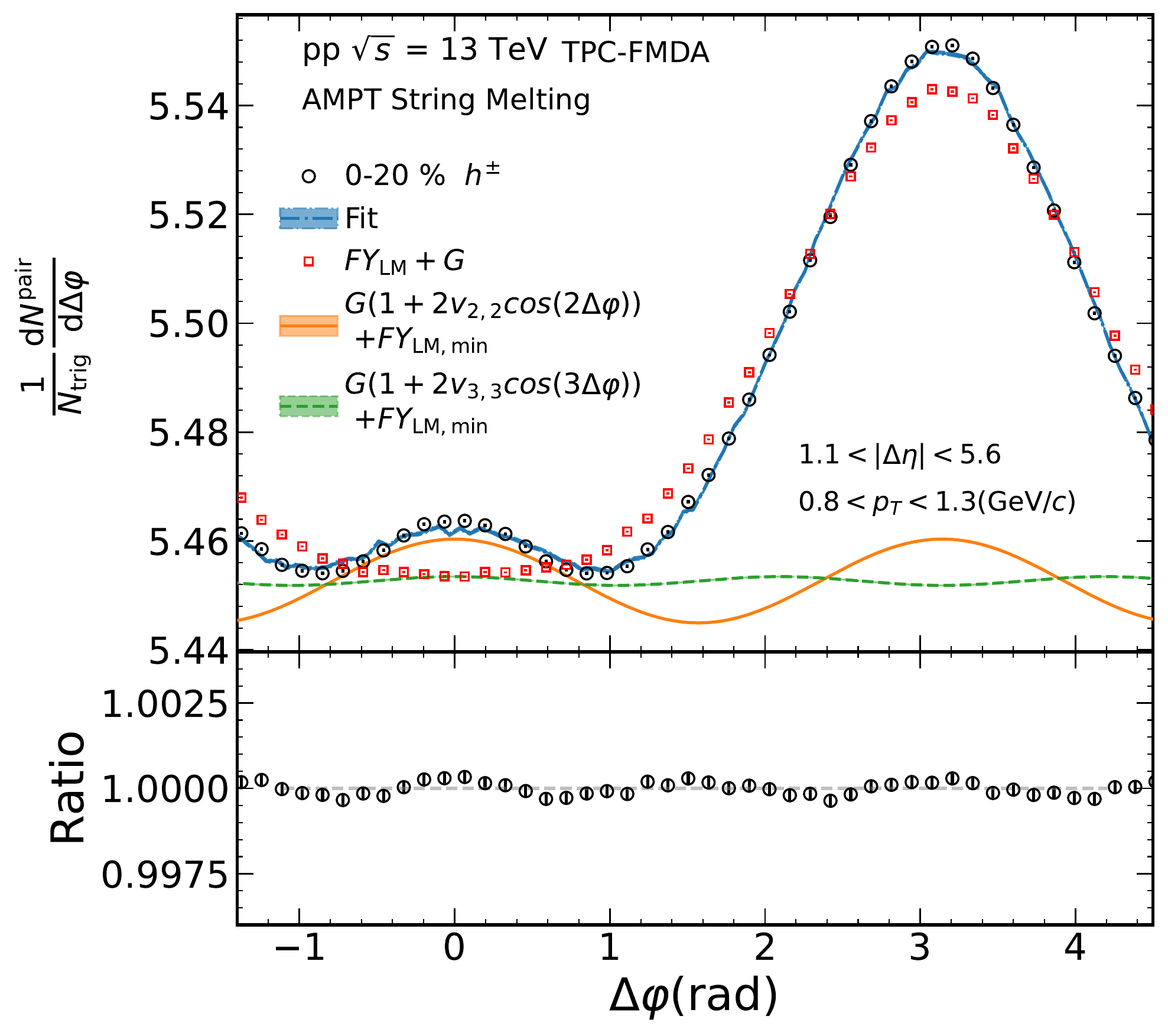} 
	\caption{The low-multiplicity template fit results. The black markers shows the signal for the 0--20\% multiplicity percentile together with its fit shown as a blue band. The red squares correspond to the low-multiplicity template. The orange and green curves correspond to the extracted $v_2$ and $v_3$ signals, respectively. (see also Fig.~\ref{fig:flowextmulti1} in Appendix from the different models.)}
	\label{fig:flowext}
\end{figure}

Fig.~\ref{fig:flowext} shows the LMTF results of TPC-FMDA correlation for 0--20\% multiplicity percentile from the AMPT String Melting configuration. 
Even with the Default $\Delta\eta$ gap, no ridge structure on the near side is seen in LM correlation function, which indicates that there is almost no jet contamination. The figure also shows the $v_{2,2}$ and $v_{3,3}$ components, yet the $v_{2,2}$ component is dominant.

\begin{figure}[tbh!]
	\centering
	\includegraphics[width=1.0\linewidth]{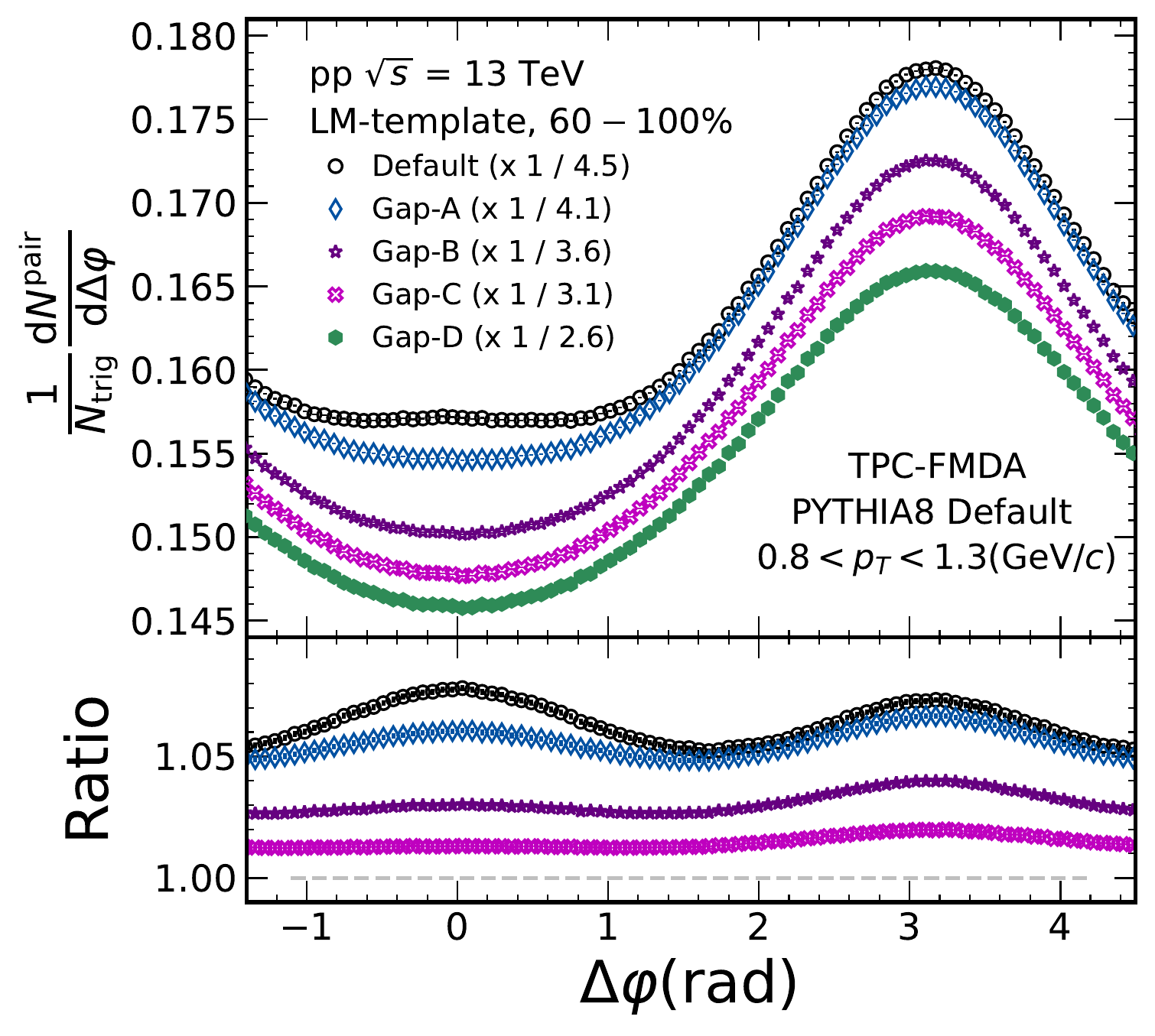} 
	\caption{The $\Delta\eta$ gap dependent LM templates with PYTHIA8 Default.}
	\label{fig:LMtemplate_pythia}
\end{figure}

The LM templates of each $\Delta\eta$ gap are seen in Fig.~\ref{fig:LMtemplate_pythia}. As the jet shape is well described in PYTHIA8 Default, the comparison is done using the PYTHIA8 model. Each template is normalised by its $\Delta\eta$. Decreasing near-side yield is seen with increasing $\Delta\eta$ gap (from Default gap to gap-D), and almost the same feature is seen in gap-C and gap-D. Under the first assumption of the template fit method, which requires no near-side yield in the low-multiplicity events, we selected the gap-D for the precise analysis. To see if the other models meet the assumption, the LM templates of each model are compared in gap-D.

\begin{figure}[tbh!]
	\centering
	\includegraphics[width=1.0\linewidth]{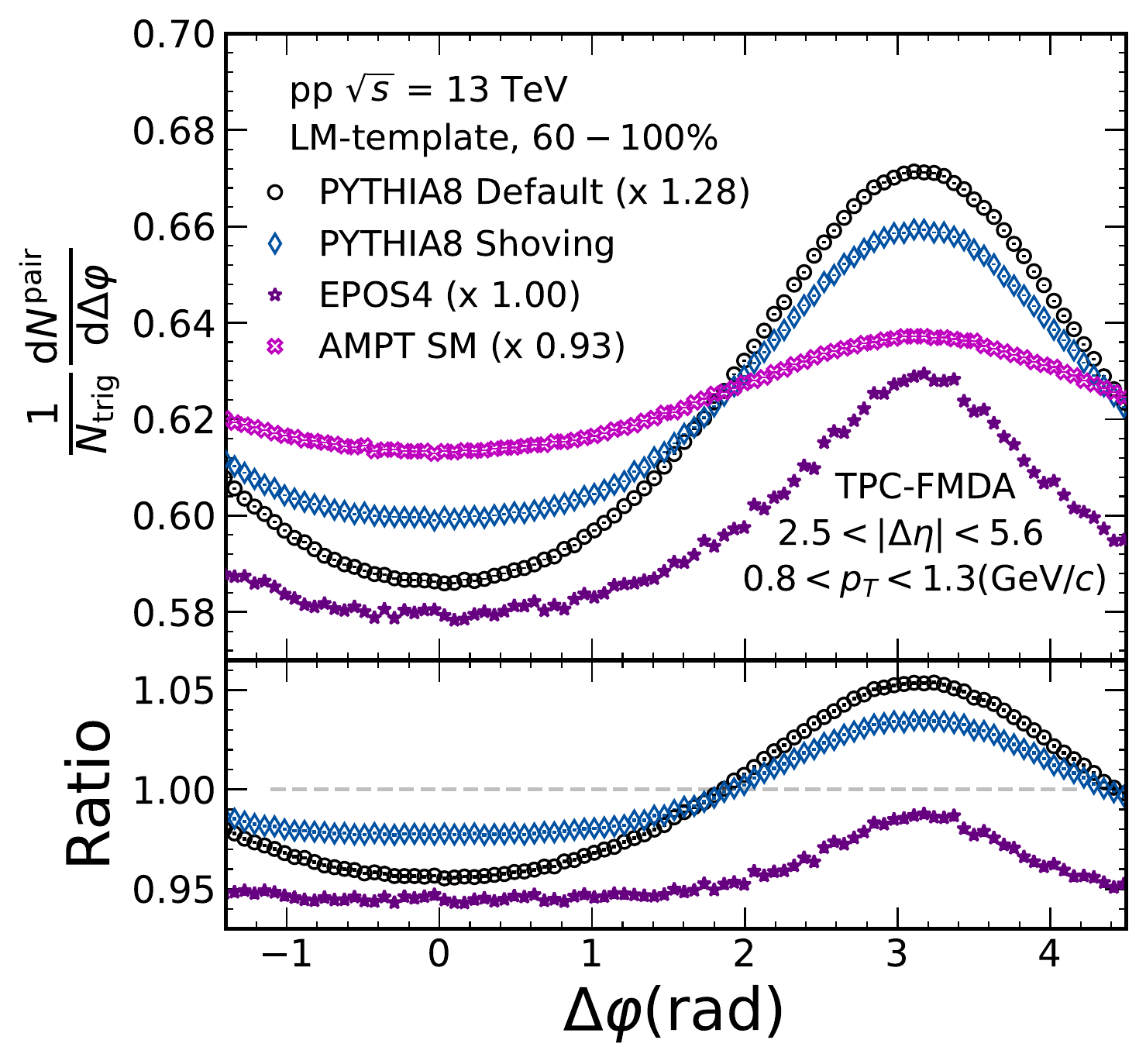} 
	\caption{The LM template for the different model calculations using the Default gap.}
	\label{fig:LMtemplate}
\end{figure}

The comparison between the LM templates of each model in the Default $\Delta\eta$ gap is seen in the Fig.~\ref{fig:LMtemplate}. As the near-side yield in the LM events comes from the jets, there should be no near-side ridge yield for the precise nonflow subtraction. The presence of the LM jet bias indicates that there is a chance of the jet shape modification in the away-side. The ratio is calculated by dividing the AMPT String Melting from the PYTHIA8 Default, PYTHIA8 String Shoving and the EPOS4 models. The PYTHIA8 shows a small near-side yield and the String Shoving shows larger yield, whilst there is no ridge yield from the AMPT String Melting and the EPOS4 models. In the case of the away-side yield, fairly broad shape is seen in the AMPT String Melting version and narrow shape in EPOS4 compared to both PYTHIA8 configurations.

However, we can not test whether the models agree with the second assumption requiring no jet shape modification depending on the event multiplicity. As every model apart from the PYTHIA8 Default contains the flow components in the away-side, we can not disentangle the flow and jets.

Finally, $v_{n}$ are extracted, based on the observed factorization of $v_{n,n}$ to single harmonics~\cite{ATLAS:2015hzw,ATLAS:2016yzd}, using the following equation,
\begin{eqnarray}
\label{eq:3}
v_{n}(p_{\rm{T,TPC}}) = \sqrt{\frac{v_{n,n}^{\rm{TPC-FMDA}} \cdot v_{n,n}^{\rm{TPC-FMDC}}}{v_{n,n}^{\rm{FMDA-FMDC}}}},
\end{eqnarray}
where $v_{n,n}(p_{\rm{T,trig}}$ and $p_{\rm{T,assoc}})$ are measured in 0.2~$<p_{\rm{T,trig}}<$~6 GeV/$c$ and integrated $p_{\rm{T}}$ ranges.

\section {Results}
\label{sec:results}
\subsection{Unidentified charged hadron flow}

The $\pt$-differential $v_{2}$ of the charged particles for different $\Delta\eta$ gap intervals in \pp\ collisions at \s\ = 13~TeV are shown in Fig.~\ref{fig:etagapdep} for several model calculations. The top panel shows the final $v_{2}$ and bottom two row panels show $v_{2,2}$ measured from TPC-FMDA and TPC-FMDC, respectively. The results for PYTHIA8 Default are shown in the first column, PYTHIA8 String Shoving in the second, EPOS4 in the third, and AMPT String Melting on the last. Even though the PYTHIA8 Default does not contain any flow component, non-zero $v_{2}$ is seen in every $\Delta\eta$ gap. As the $\Delta\eta$ gap becomes larger, the less non-flow dominant region we contain as shown in the Fig.~\ref{fig:2DPythiaDefault}, therefore smaller amplitude of $v_{2}$ is seen with increasing $\Delta\eta$ gap. Despite having both flow and non-flow components in the PYTHIA8 String Shoving, similar behaviour is seen with the PYTHIA8 Default with smaller magnitude of the flow component in overall. This can be due to the presence of the near-side yield in the low multiplicity which can be seen in the template fit results (see Fig.~\ref{fig:flowextmulti1} in Appendix). 
In the case of the EPOS4, which also includes the flow components, smaller magnitude of $v_{2}$ and $v_{2,2}$ are seen compared to the both PYTHIA8 configurations and similar $\pt$ and $\Delta\eta$ gap dependence is seen with PYTHIA8. Lastly, the AMPT String Melting model shows that in low \pt{} regions $v_{2}$ doesn't vary much on the $\Delta\eta$ gap selection. However, the $v_{2}$ increases with increasing $\Delta\eta$ gap unlike other models and these are mostly influenced by the fact that the TPC-FMDC has effected by the jet contamination in smaller $\Delta\eta$ gap selections as seen in the bottom panel of AMPT (see the correlation function). 
In the low \pt{} regions, $v_{2}$ are increased by 50 $\%$ and in high \pt{} regions, a factor of two respectively (see Fig.~\ref{fig:ampt_nonflow} in Appendix). Since the largest $\Delta\eta$ gap has the smallest contribution from non-flow, in latter sections, only results from the AMPT SM with the gap-D will be shown.
\begin{widetext}
    \begin{minipage}{\linewidth}
    \begin{figure}[H]
    \centering
		\includegraphics[width=0.9\textwidth]{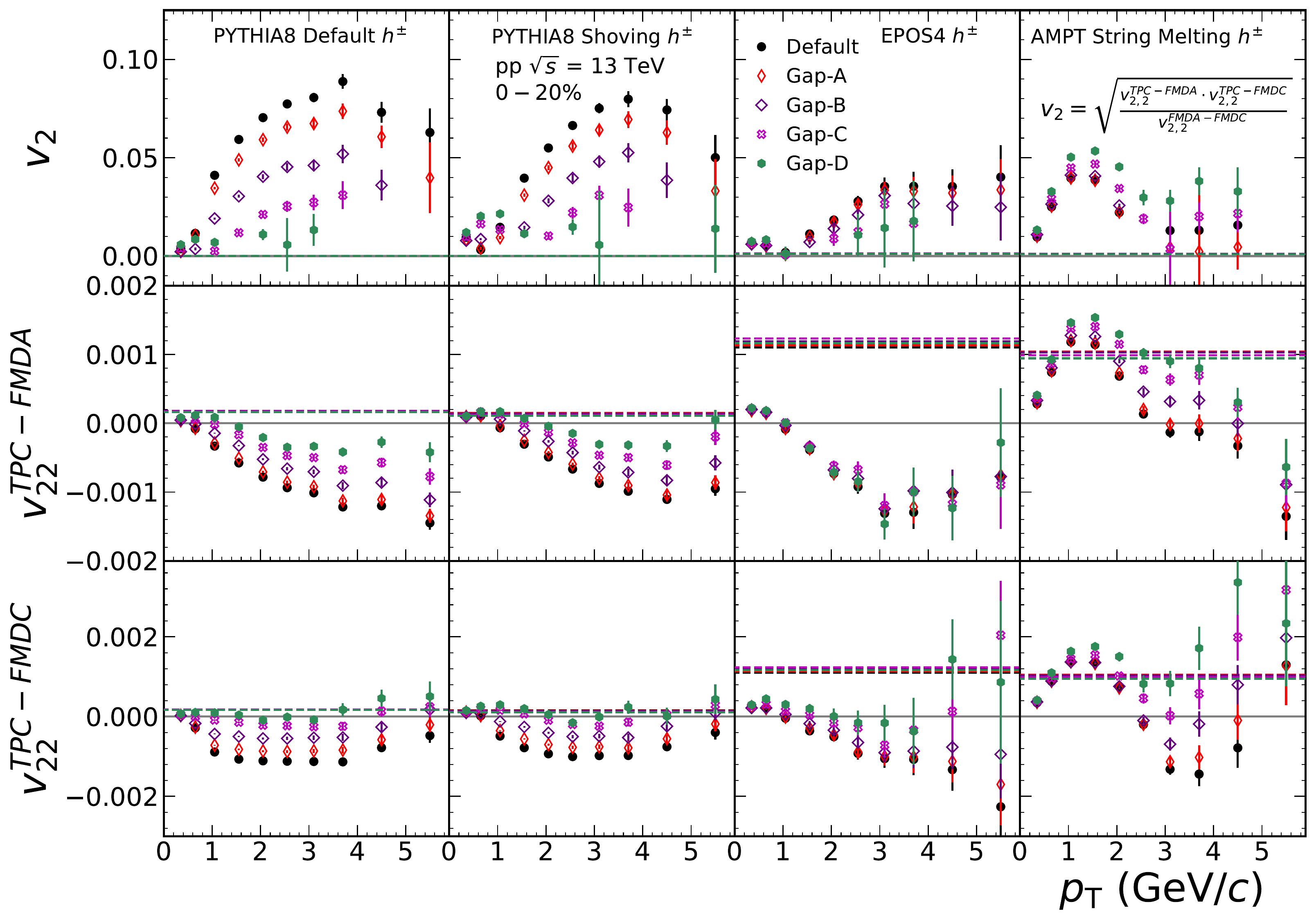} 
\caption{The $\pt$-differential $v_{2}$ for different $\Delta\eta$ gap intervals at \s\ = 13~TeV for several model calculations are shown for the charged particles. Two components to calculate the final $v_2$ on the top are shown in the bottom two panels. The results of $v_{2,2}^{\rm{FMDA-FMDC}}$ and $v_{2}^{\rm{FMDA-FMDC}}$
\label{fig:etagapdep} are shown as dashed lines on each panel.}
    \end{figure}
    \end{minipage}
    \clearpage
\end{widetext}

\begin{widetext}
    \begin{minipage}{\linewidth}
    \begin{figure}[H]
      \centering
		\includegraphics[width=0.675\textwidth]{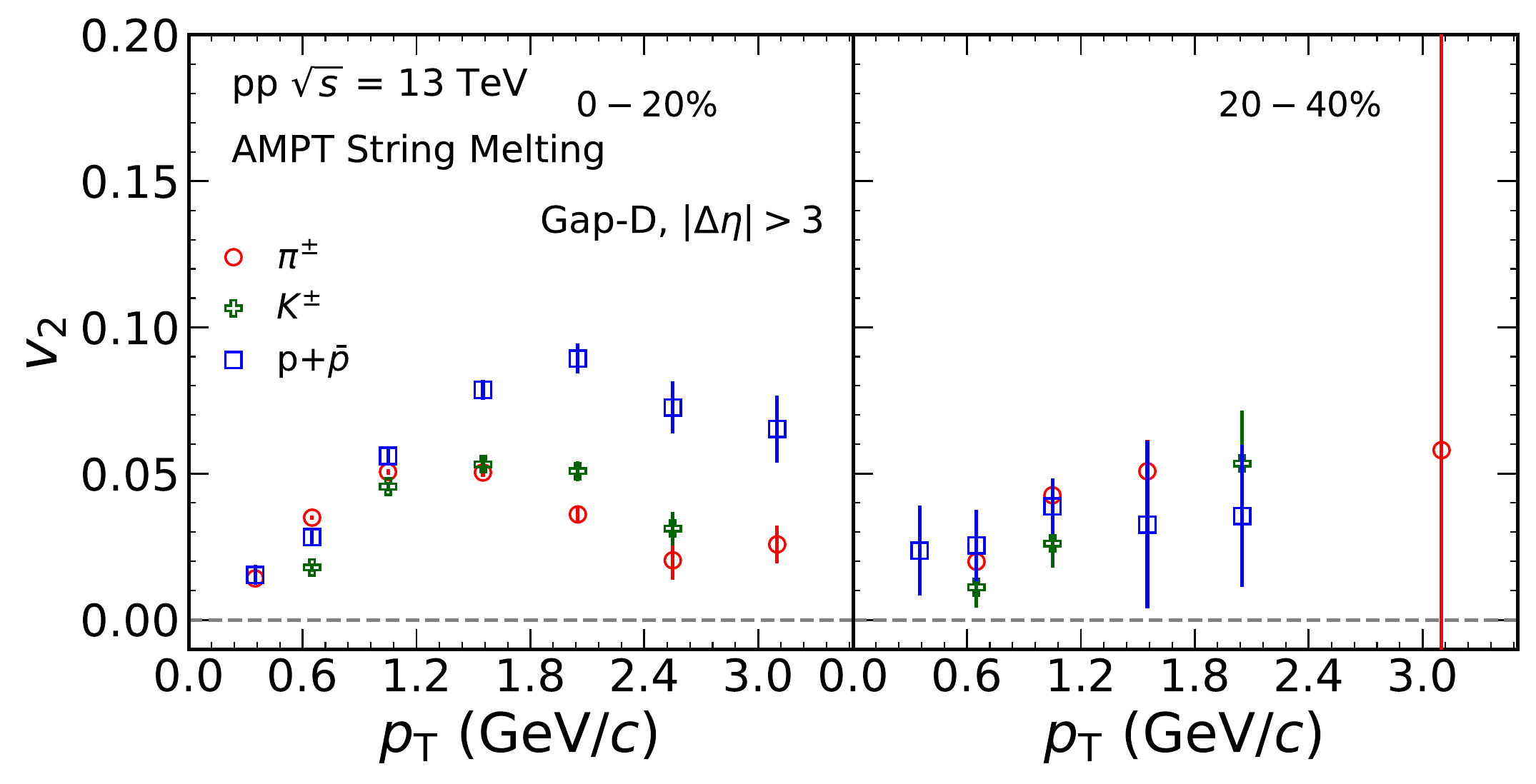} 
\caption{The $\pt$-differential $v_{2}$ for different particle species in 0--20\% and 20--40\% multiplicity percentiles in \pp\ collisions at \s\ = 13~TeV from the AMPT String Melting model calculations. }
\label{fig:ampt_pid_pt}
    \end{figure}
    \end{minipage}
\end{widetext}
\subsection{Identified charged hadron flow}

Fig.~\ref{fig:ampt_pid_pt} shows the $v_{2}$ of the identified charged particles in 0--20\% and 20--40\% events with the AMPT String Melting model. Grouping of $v_{2}$ is seen depending on the particle species, especially whether the particle is meson or baryon in 0--20\% events. In the case of the 20--40\% events, the mass splitting is not clearly seen mostly due to the lack of the statistics. Also, as the smaller $v_{2}$ is seen in 20--40\% compared to 0--20\%, we also studied about the multiplicity dependence of $v_{2}$.

Fig.~\ref{fig:ampt_pid_mt} shows the dependence of $v_{2}$ on transverse kinetic energy, normalized by the number of quark constituents ($n_q$), using the AMPT String Melting model. The transverse kinetic energy, ${\rm KE}_{\rm{T}}$, is defined as ${\rm KE}_{\rm{T}} = m_{\rm{T}} - m_{\rm{0}}$, where $m_{\rm{T}}= \sqrt{m_{0}^2 + p_{\rm{T}}^2}$ is the transverse mass.

The observed variation of flow with particle species arises from the hydrodynamic pressure gradient, which is dependent on the particles' mass. As ${\rm KE}_{\rm{T}}$ is directly related to the pressure gradient, we measured the transverse kinetic energy-dependent flow. We normalized the $v_{2}$ and ${\rm KE}_{\rm{T}}$ by the number of quark constituents, as the number of quarks in a particle varies by its type.
While previous data from large collision systems at LHC show that the flow coefficients approximately lie on a line regardless of the particle species~\cite{ALICE:2014wao,ALICE:2015lib,ALICE:2016cti,ALICE:2018yph}, the AMPT results in \pp\ collisions show some deviation from the scaling in both 0--20\% (left) and 0--5\% (right) events. Experimental results obtained with the LMTF method over a wider range of $p_{\mathrm{T}}$ will provide further insight into the presence of NCQ scaling in small system collisions.

\begin{widetext}
    \begin{minipage}{\linewidth}
    \begin{figure}[H]
    \centering
		\includegraphics[width=0.675\textwidth]{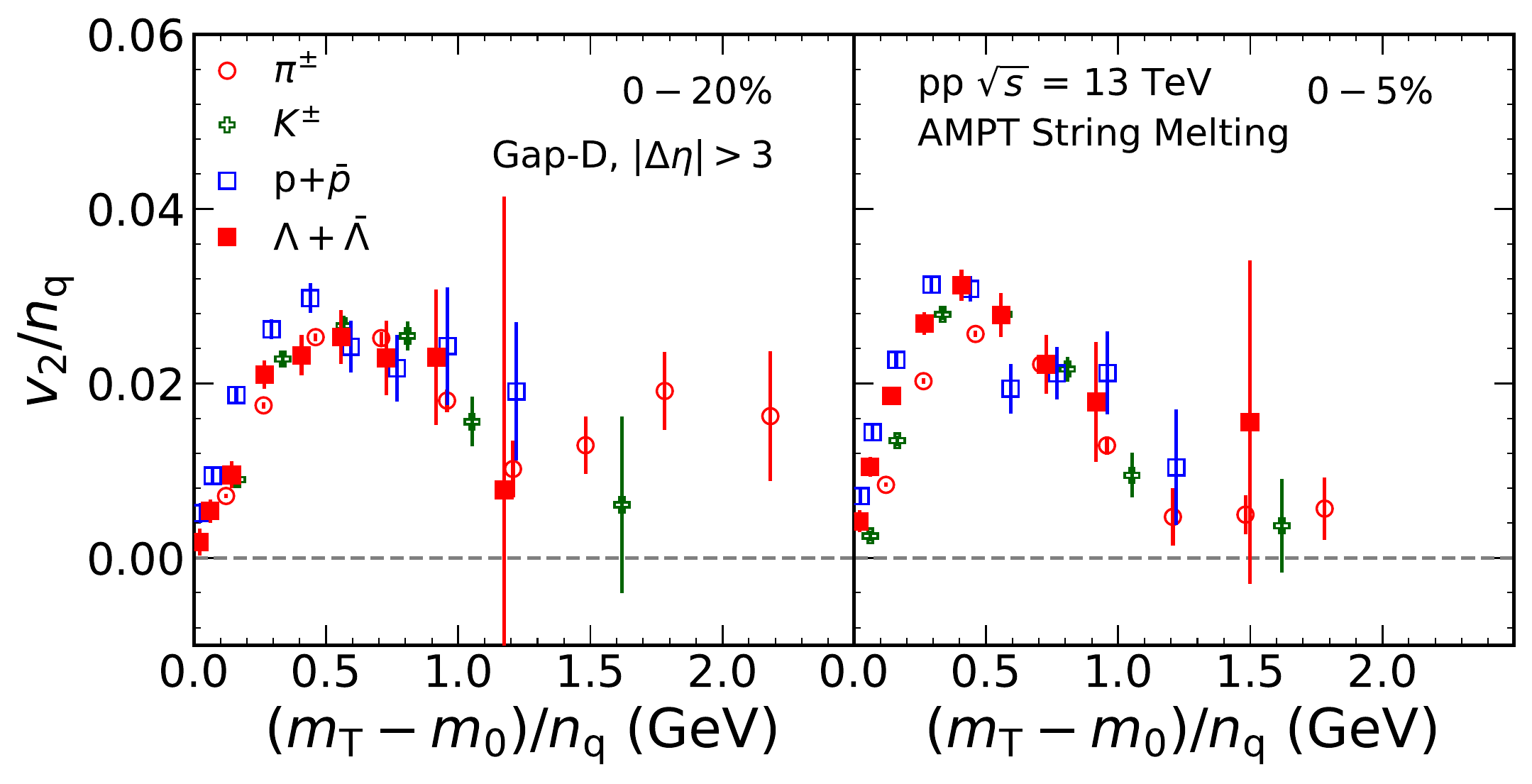} 
\caption{The NCQ scaled $\mt$-dependent $v_{2}$ for different particle species in 0--20\% (left) and 0--5\% (right) high multiplicity percentiles in \pp\ collisions at \s\ = 13~TeV from the AMPT String Melting model calculations. }
\label{fig:ampt_pid_mt}
    \end{figure}
    \end{minipage}
\end{widetext}

\subsection{Multiplicity dependent flow}

\begin{figure}[ht!]
		\includegraphics[width=\columnwidth]{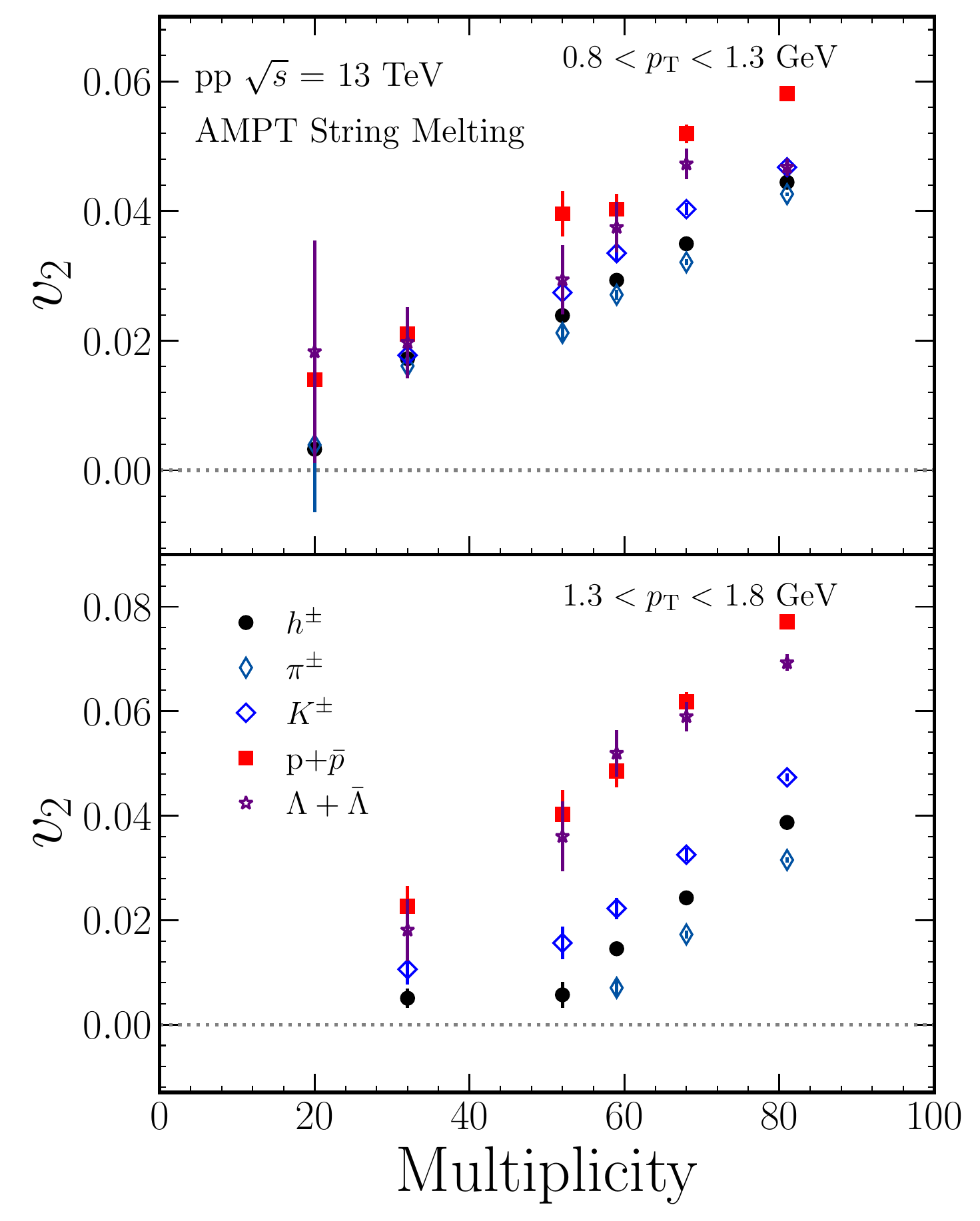}
    \caption{The multiplicity dependence of $v_2$ for different particle species in \pp\ collisions at \s\ = 13~TeV from the AMPT String Melting model calculations.}
    \label{fig:pid_mult_dep_2bins}
\end{figure}

In Fig.~\ref{fig:pid_mult_dep_2bins}, we present the magnitude of $v_2$ as a function of multiplicity for various particle species in two $p_\mathrm{T}$ ranges. The $|\Delta\eta|$ range considered is $>3$, and $v_2$ is shown for $0.8<p_{\mathrm{T}}<1.3,\mathrm{GeV}/c$ and $1.3<p_{\mathrm{T}}<1.8,\mathrm{GeV}/c$. Firstly, we observe that the magnitude of $v_2$ increases with increasing multiplicity for both $p_\mathrm{T}$ ranges, regardless of the particle type. Secondly, $v_2$ decreases towards lower multiplicities and starts to saturate at a multiplicity of around 50. While the AMPT String Melting model shows a linear multiplicity dependence, the experimental results reported in Refs.~\cite{ATLAS:2015hzw,ATLAS:2016yzd, Khachatryan:2015lva} show a mild decrease towards low multiplicity events.

In the case of the higher $p_\mathrm{T}$ range shown in the bottom panel of Fig.~\ref{fig:pid_mult_dep_2bins}, we observe that the multiplicity dependence of charged hadrons differs from that of identified mesons in the first two multiplicity bins. Interestingly, baryons do not show this saturation yet in those multiplicity ranges, within the uncertainties. Furthermore, the ordering in the $v_2$ magnitudes between different particle species is visible, as discussed in the previous section. For both $\pt$ ranges, the magnitudes of $v_2$ are clearly separated between mesons and baryons in higher multiplicities.

 \section {Conclusions}
\label{sec:summary}
We extracted flow coefficients for various particle species, including $\pi$, $K$, $p$, and $\Lambda$, with identified hadrons using multiple MC generators and detector combinations in wide $\Delta\eta$ ranges for \pp\ collisions at \s\ = 13 TeV. The flow measurements were obtained through long-range correlations in different high-multiplicity classes by employing the LMTF method. This approach enabled us to eliminate the enhanced away-side jet fragments in high-multiplicity events relative to low-multiplicity events. However, we found that subtracting non-flow contamination in small systems could lead to biased results, due to the kinematic bias on jets and different model implementations of flow and jet components. Specifically, we observed that the PYTHIA8 Default model, which does not account for collective flow, produces biased results towards large flow. Moreover, it was not possible to extract flow signals from the EPOS4 and PYTHIA8 Shoving models, which contain flow components, as they violate the assumptions of the template fit method, containing near-side yield in low-multiplicity events.
We conducted studies of the LM-template method in multiple $\Delta\eta$-gaps and found that the current ALICE $\eta$ acceptance might still be influenced by non-flow contamination, suggesting the need for larger $\Delta\eta$-gaps in future analyses. Only the AMPT String Melting model among the studied models was free from this bias and showed a mass ordering at low $p_{\mathrm{T}}$ and particle type grouping in the intermediate $p_{\mathrm{T}}$ range, similar to what is observed in large systems. However, this ordering was quite distinct from that seen in large systems.

\section*{Acknowledgments}
We thank Klaus Werner, Christian Bierlich and Zi-Wei Lin for fruitful discussions with their model calculations. 
We acknowledge CSC - IT Center for Science in Espoo, Finland, for the allocation of the computational resources. 
MV, TK, and DJK are supported by the Academy of Finland, the Centre of Excellence in Quark Matter (project 346328).
SJ and SHL are supported by the National Research Foundation of Korea (NRF) grant funded by the Korea government (MSIT) under Contract No. 2020R1C1C1004985. We also acknowledge technical support from KIAF administrators at KISTI.

\nocite{*}
\bibliography{paper}
\clearpage
\appendix*
\section{Supplemental Material}
\label{sec:supplemental}
In this section, the low multiplicity template fits and two-particle correlation functions of the models used in the paper are described in further detail. Also few additional figures are provided for the $\eta$ gap and multiplicity dependence flow results that are from the AMPT String Melting model calculations.

\subsection{Low multiplicity template fit}
 Fig.~\ref{fig:flowextmulti1} shows the LM-template fit results for pp $\sqrt{s} =$ 13 TeV TPC-FMDA correlations for three models in the $0-20\%$ multiplicity percentiles. In all columns, the per-trigger particle yield is shown as a function of $\Delta\varphi$. The black markers represent HM events, the blue bands represent the fit, and the red squares represent LM events. Orange and green bands represent the extracted harmonic flow components. The bottom panels show a zoomed-in view of the near-side region to better visualize the data.

The analysis done to the EPOS4 model calculations result relatively flat near-side LM yields, while both PYTHIA8 models show distinct near-side ridge. Notably, PYTHIA8 Shoving results as largest near-side yields in the LM events, which are comparable to the yields in the HM events. Additionally, in Fig.~\ref{fig:flowextmulti1}, the PYTHIA8 models have positive $v_3$ and negative $v_2$, while the EPOS4 model has positive $v_3$ and $v_2$.

 \begin{widetext}
    \begin{minipage}{\linewidth}
    \vspace{1cm}
    \begin{figure}[H]
        \centering
		\includegraphics[width=\textwidth]{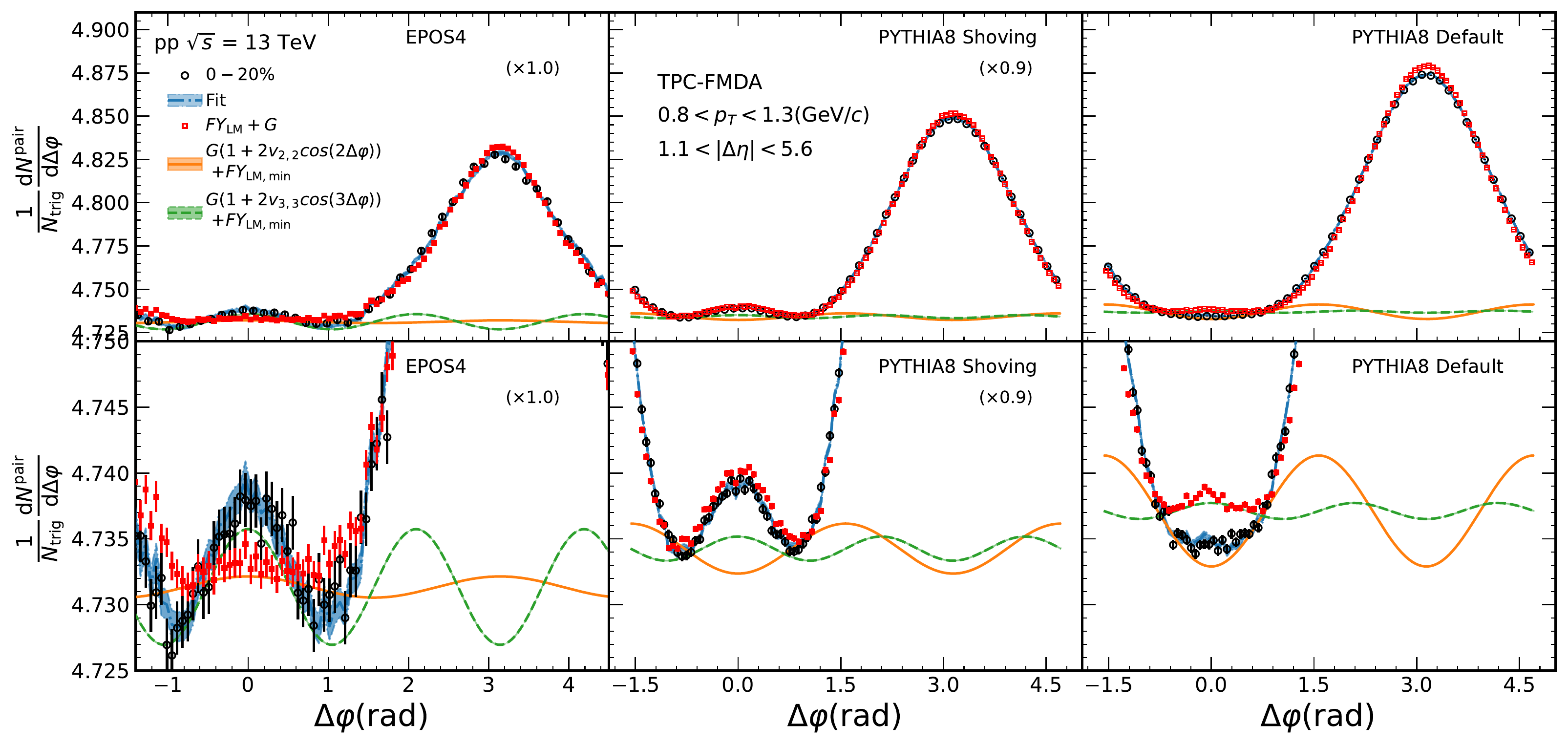} 
	\caption{The template fit results with the LM-templates from EPOS4, PYTHIA8 Shoving and PYTHIA8 Default models. The black markers shows the signal for the $0-20\%$ multiplicity percentile together with its fit shown as a blue band. The red squares correspond to the LM signal. The orange and green curves correspond to the extracted $v_2$ and $v_3$ signals, respectively.}
	\label{fig:flowextmulti1}
    \end{figure}
    \end{minipage}
    \clearpage
\end{widetext}

 \begin{widetext}
    \begin{minipage}{\linewidth}
    \begin{figure}[H]
        \centering
		\includegraphics[width=0.8\textwidth]{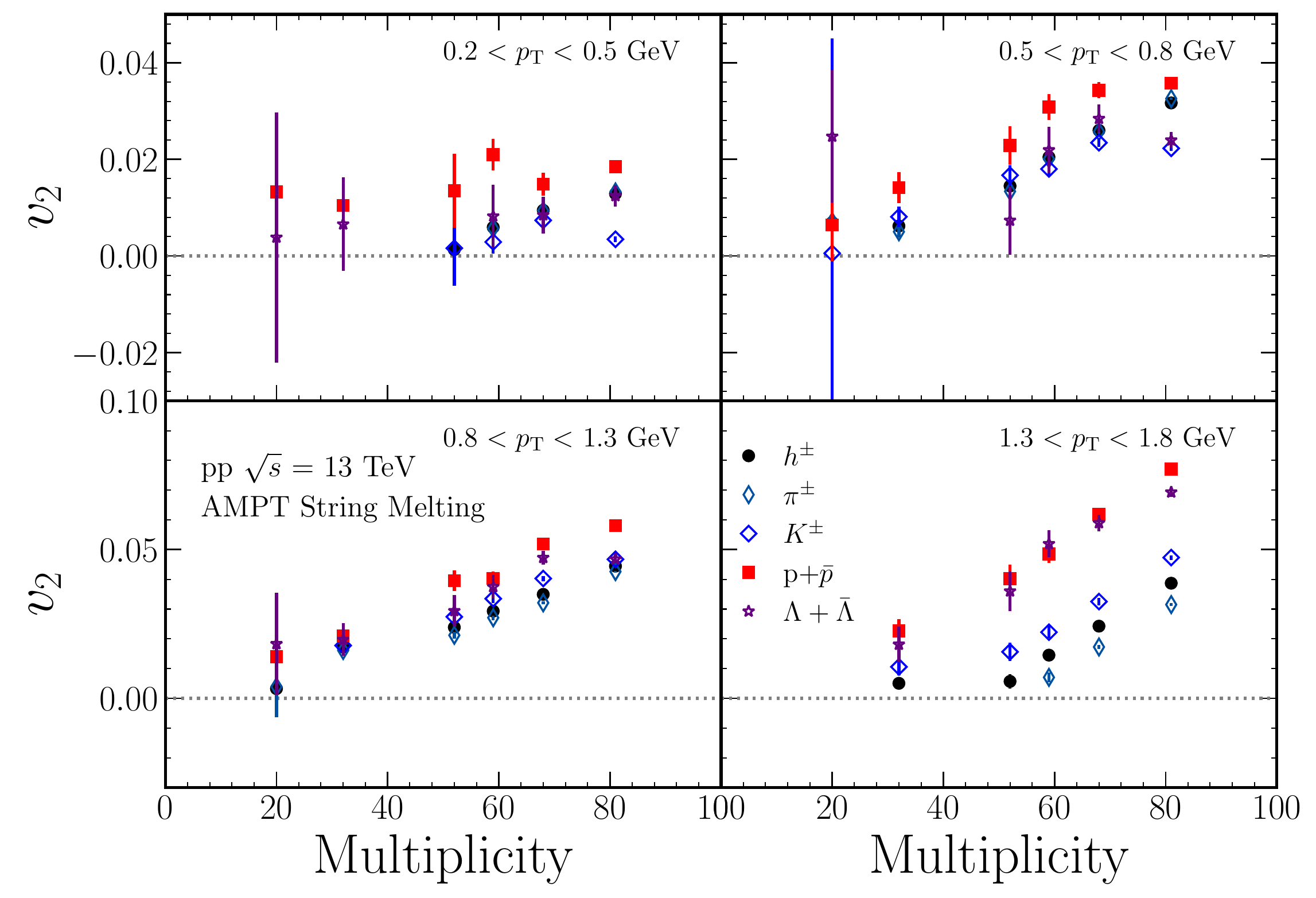}
    \caption{The multiplicity dependence of $v_2$ with various \pt\ bins for different particle species in pp collisions at \s\ = 13~TeV from the AMPT String Melting model calculations.}
    \label{fig:pid_mult_dep}
    \end{figure}
    \end{minipage}
\end{widetext}

\subsection{Multiplicity and $\eta$ gap dependence}

In Fig. \ref{fig:pid_mult_dep}, we present the magnitude of $v_2$ as a function of multiplicity for various particle species in four $p_\mathrm{T}$ ranges as an addition to sec.~\ref{sec:results}, where only two high $p_\mathrm{T}$ ranges was provided. The conclusions are same for $0.5<p_{\mathrm{T}}<0.8,\mathrm{GeV}/c$ as they were for $0.8<p_{\mathrm{T}}<1.3,\mathrm{GeV}/c$ and $1.3<p_{\mathrm{T}}<1.8,\mathrm{GeV}/c$.
In the case of the lowest $p_\mathrm{T}$ range shown in the top left panel of Fig.~\ref{fig:pid_mult_dep}, we observe that the multiplicity dependence is weaker than with other $p_\mathrm{T}$ ranges within the uncertainties. Interestingly, baryons do not show this saturation yet in those multiplicity ranges, within the uncertainties. Furthermore, the ordering in the $v_2$ magnitudes between different particle species is visible, as discussed in the previous section. At higher $\pt$ ranges, the magnitudes of $v_2$ are clearly separated between mesons and baryons in higher multiplicities.

\begin{figure}[ht!]
\centering
    \includegraphics[width = 0.49\textwidth]{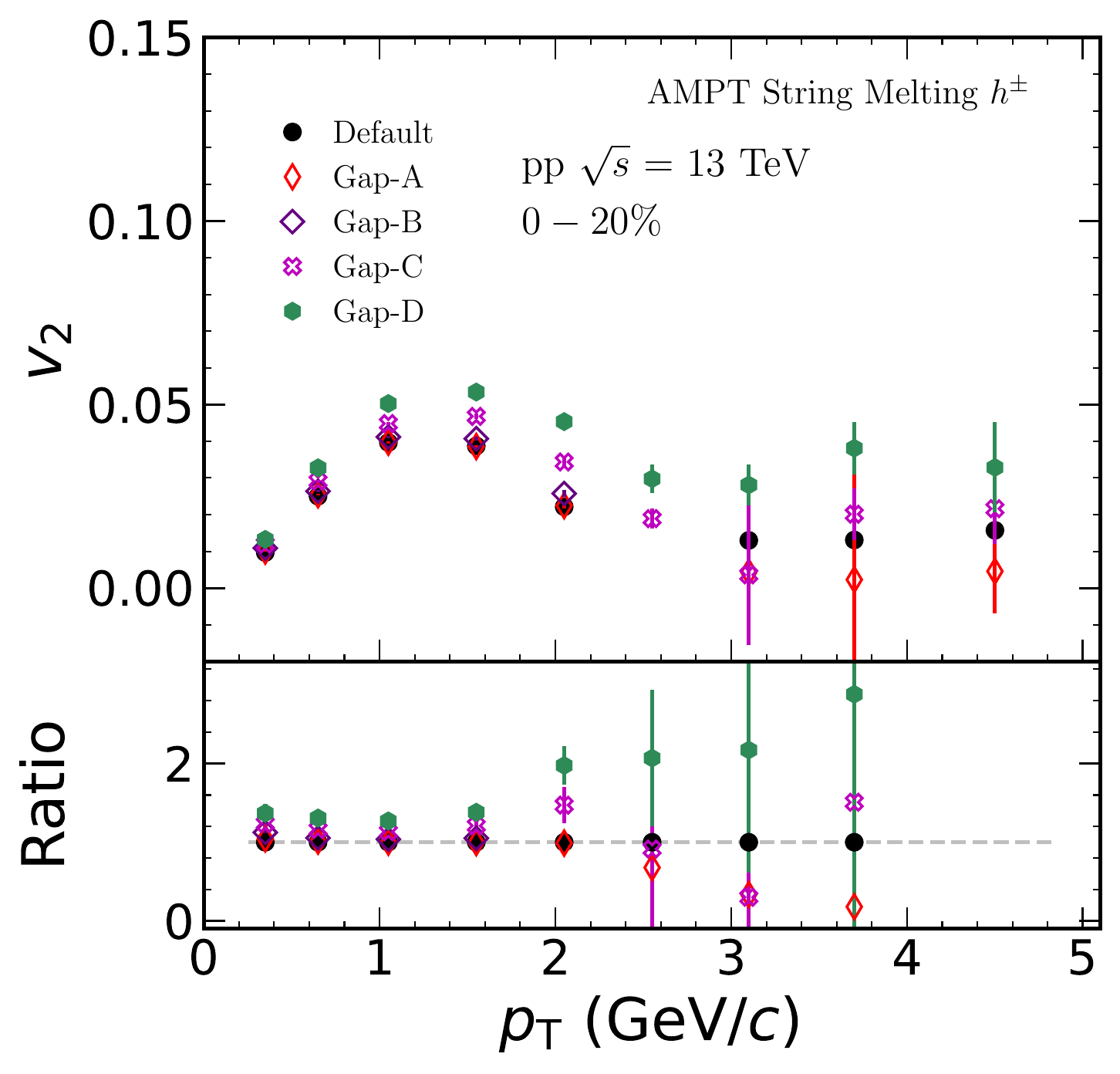}
    \caption{The $\pt$-differential $v_{2}$ for different $\eta$ gap interval at \s\ = 13~TeV for AMPT String Melting calculations are shown for the charged particles.}
    \label{fig:ampt_nonflow}
\end{figure}

The panel for the AMPT String melting model in Fig.~\ref{fig:etagapdep} is zoomed in Fig.~\ref{fig:ampt_nonflow} and the ratio of the results from the different  $\eta$ gap to the default selection is shown in the bottom panel.
In the low \pt{} regions, $v_{2}$ are increased by 50 $\%$ and in high \pt{} regions, a factor of two respectively.

\subsection{Two dimentional correlation functions}
\begin{figure*}[ht!]
  \centering
		\includegraphics[width=\textwidth]{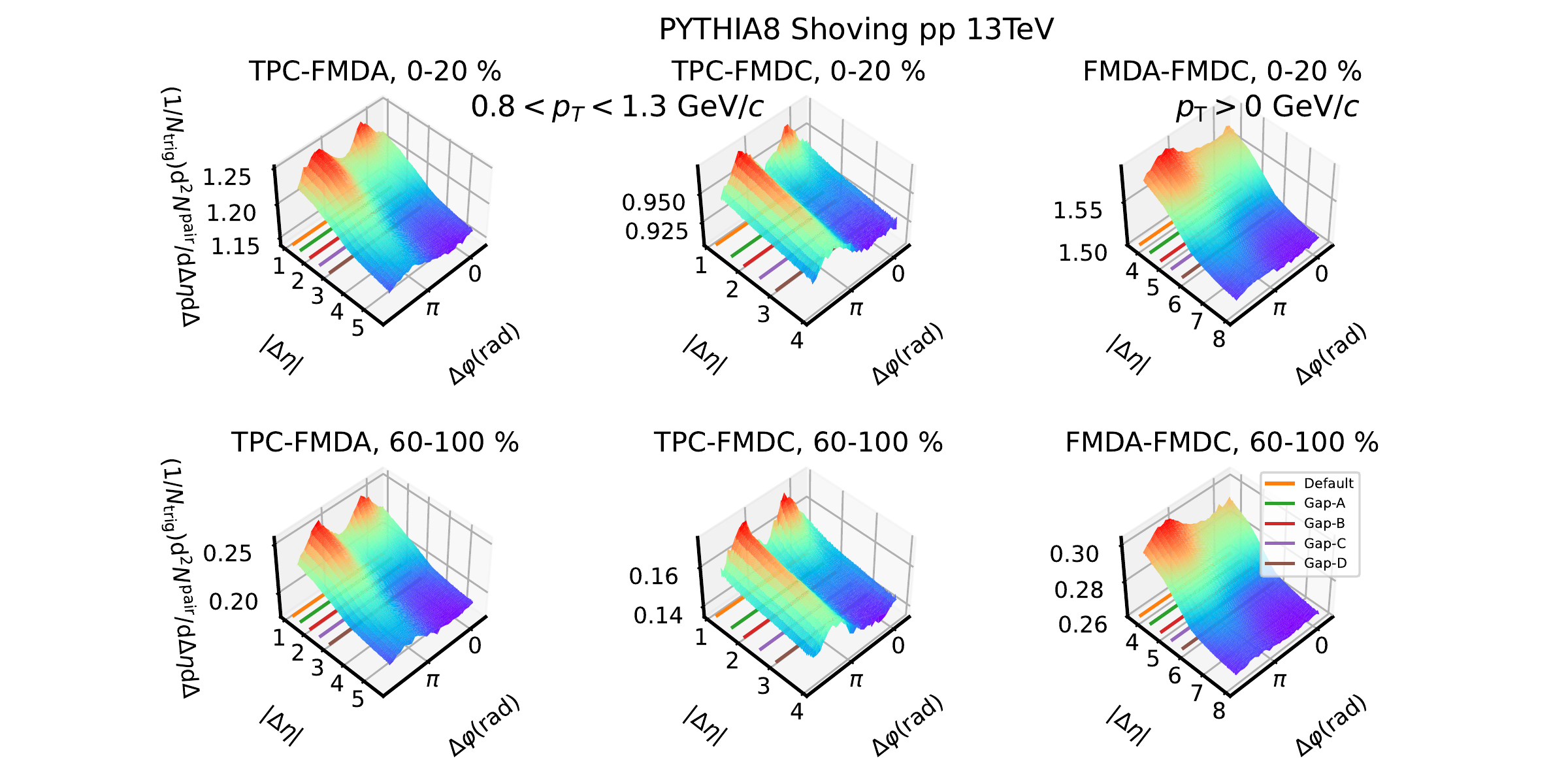} 
	\caption{Two-particle correlation functions as functions of $\Delta\eta$ and $\Delta\varphi$ for HM(0--20\%, top panels) and LM(60--100\%, bottom panels) events by using various combinations of the detectors in $\sqrt{s}=13$ pp collisions from PYTHIA8 String Shoving calculations. The intervals of $\pttrig$ and $\ptassoc$ are 0.8~$<\it{p}_{\rm{T}}<$~1.3~GeV/$c$ with TPC and $\it{p}_{\rm{T}}>$~0~GeV/$c$ with FMDA or FMDC.}
	\label{fig:flowextmulti2}
\end{figure*}

\begin{figure*}[ht!]
  \centering
		\includegraphics[width=\textwidth]{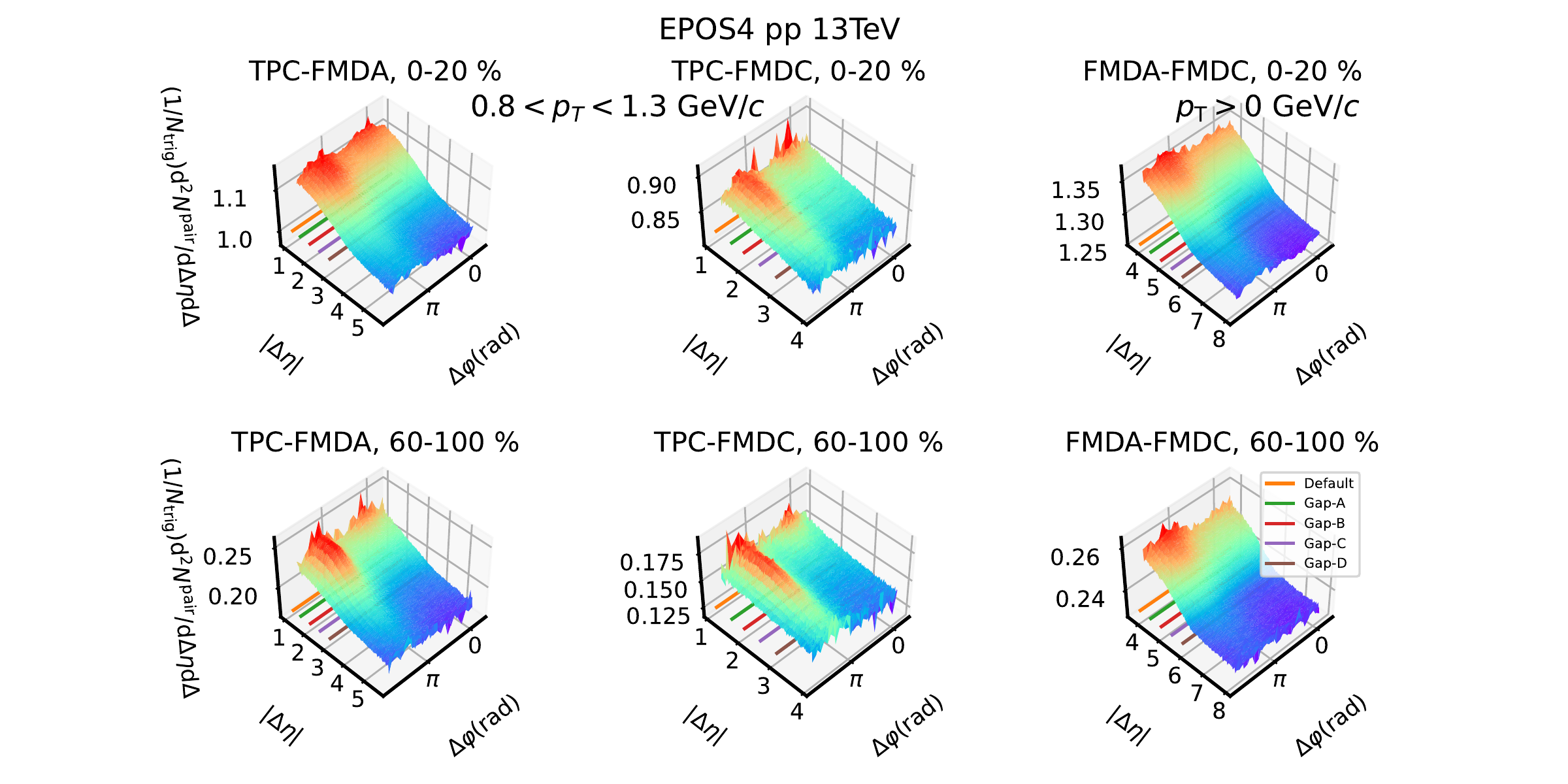} 
	\caption{Two-particle correlation functions as functions of $\Delta\eta$ and $\Delta\varphi$ for HM(0--20\%, top panels) and LM(60--100\%, bottom panels) events by using various combinations of the detectors in $\sqrt{s}=13$ pp collisions from EPOS4 calculations. The intervals of $\pttrig$ and $\ptassoc$ are 0.8~$<\it{p}_{\rm{T}}<$~1.3~GeV/$c$ with TPC and $\it{p}_{\rm{T}}>$~0~GeV/$c$ with FMDA or FMDC.}
	\label{fig:flowextmulti3}
\end{figure*}

 Fig.~\ref{fig:flowextmulti2} and Fig.~\ref{fig:flowextmulti3} show two-particle correlation functions from PYTHIA8 Shoving and EPOS4 in two different multiplicity percentiles ($0-20\%$ and $60-100\%$). The figures are divided into six panels, each panel showing correlation functions from separate detector combinations. The $\Delta\eta$ gaps used in our analysis are expressed as colored lines parallel to $\Delta\varphi$ line marking the minimum value of each gap. 
 In the lower $\Delta\eta$ region, TPC-FMDA and TPC-FMDC results are showing clear peaks. For TPC-FMDA and FMDA-FMDC, a noticeable decrease in the amount of correlations can be seen when going into larger $\Delta\eta$ regions relative to TPC-FMDC. This is also seen in the other models as well (see Fig. \ref{fig:2DAMPT_SM} and Fig. \ref{fig:2DPythiaDefault}). The correlations from FMDA-FMDC results are showing larger without a clear peak. For TPC-FMDC, the away-side yields are elongated along the whole $|\Delta\eta|$ range. 

\end{document}